# Principal Component–Based Estimation of Finite Population Mean under Multicollinearity

**Rajesh Singh[1], Shobh Nath Tiwari[1, *]**

[1]Department of Statistics, Institute of Science, Banaras Hindu University, Varanasi-221005, Uttar Pradesh, India

Emails:rsinghstat@bhu.ac.in;shobh7285@gmail.com

[*]**Corresponding:**shobh7285@gmail.com

**Abstract**

Auxiliary information is frequently utilized in survey sampling to improve the efficiency of estimators of the finite population mean. However, the simultaneous use of multiple auxiliary variables often induces multicollinearity, which adversely affects the stability and performance of conventional estimators. To address this issue, the present study proposes a principal component analysis (PCA)–based estimation approach for the finite population mean in the presence of multicollinearity between two auxiliary variables. The proposed methodology transforms the correlated auxiliary variables into a set of orthogonal principal components, thereby removing the effect of multicollinearity while preserving the essential information contained in the auxiliary variables. An efficient estimator is then constructed using these components under simple random sampling without replacement. The bias and mean square error (MSE) of the proposed estimator are derived up to the first order of approximation. The performance of the proposed estimator is evaluated through both empirical and simulation studies under varying correlation structures. Moreover, the presence of multicollinearity is evaluated using variance inflation factors, condition indices, and eigenvalues. The results from empirical and simulation studies demonstrate that the proposed PCA-based estimator outperforms several conventional estimators in terms of MSE and percentage relative efficiency (PRE) when multicollinearity exists, ensuring robust and efficient estimation of the population mean.

## 1. Introduction

In sample survey methodology, the incorporation of auxiliary information at the design stage, estimation stage, or both, has been widely recognized as an effective approach for improving the precision of population parameter estimates.Notably, the estimation of the population mean remains a primary objective, as it offers a concise and informative representation of the distribution of the study variable.However, the level of accuracy achieved is influenced by various factors, including sample size, population heterogeneity, and the presence of auxiliary information.In the existing literature, numerous estimators based on auxiliary information, including ratio, product, and regression-type estimators, have been developed to improve estimation performance.In particular, ratio and regression estimators are widely recognized for their ability to lower the MSE by incorporating auxiliary information (Bahl and Tuteja[1]). The use of auxiliary information strengthens the estimation procedure by leveraging the underlying statistical relationship between the study variable and related supplementary information (Isaki and Fuller[2]; Särndal et al.[3]). A considerable body of research has focused on estimating the population mean using simple random sampling (SRS), resulting in a wide range of estimators proposed in key studies including (Kadilar and Cingi[4]; Singh et al.[5]; Sajjad and Ismail[6]; Shabbir and Gupta[7]; Singh et al.[8]; Upadhyaya and Singh[9]).Recent research has contributed significantly to the development of population mean estimation, with notable studies including Singh et al. [10]; Singh and Tiwari [11];Singh et al. [12];Singh et al. [13], and other related literature.

The integration of multiple auxiliary variables into survey sampling frameworks plays a crucial role in enhancing the accuracy and efficiency of population parameter estimation. Raj [14]highlighted the role of multiple auxiliary variables in variance reduction. Ahmad et al. [15] developed a generalized estimator of the population mean for two-phase sampling with multiple auxiliary variables in the presence of first-phase nonresponse and absence of prior information.Sher et al. [16] developed enhanced estimators of the population mean in SRS using two auxiliary variables and demonstrated their practical utility. Almulhim et al. [17] addressed the estimation of the population mean with dual auxiliary information under non-response using SRS.However, the use of several auxiliary variables may induce multicollinearity, a condition that magnifies standard errors, undermines estimator performance, and limits the interpretability of individual variable effects (Kmenta and Klein [18]; Gujarati[19]). Following Jolliffe [20], this study introduces and validates a PCA-based estimator using simulation and real data, demonstrating its improved efficiency over conventional regression estimators when auxiliary variables are correlated.

In the context of survey sampling with multiple auxiliary variables, it is generally assumed that these variables are not highly correlated. However, in practical situations, auxiliary variables often exhibit strong intercorrelations, leading to the problem of multicollinearity.The presence of multicollinearity adversely affects the accuracy and stability of conventional estimators, thereby limiting their performance in practical scenarios with highly correlated predictors.To address this issue, PCA has emerged as an effective tool for transforming correlated auxiliary variables into uncorrelated components while preserving essential variability. Motivated by this, the present study develops a log-type estimator based on principal components for estimating the population mean under multicollinearity. In practical survey settings, auxiliary variables often exhibit skewness, nonlinear associations, and large fluctuations, which not only induce multicollinearity but also compromise the stability of conventional estimators. While PCA effectively resolves the issue of multicollinearity through orthogonal transformation, it does not fully address the effects of skewness and extreme variations. Therefore, the incorporation of log-type transformation further enhances estimator performance by stabilising extreme ratios and reducing variance, particularly when auxiliary variables deviate substantially from their population values.The primary objective of this study is to offer methodological guidance for the application of PCA-based approaches in practical survey sampling settings, particularly where accuracy, stability, and efficiency are essential.

## 2. Notation and Definitions

Let there be a finite population $P = (p_1, p_2, \ldots, p_N)$ of $N$ identifiable units, from which a subset of $n$ elements is randomly drawn using the SRSWOR procedure.Consider $X_i$ and $Z_i$ as auxiliary variables and $Y_i$ as the study variable associated with the $i^{th}$ unit, where $i = 1, 2, \ldots, N$.Similarly, let $x_i$, $z_i$, and $y_i$ represent the values associated with the $i^{th}$ unit in the sample $(i = 1, 2, \ldots, n)$.The parameters are specified as follows.

**Table 1:**Notations and their descriptions.

| Notation | Description |
|---|---|
| $N$ | Population size. |
| $n$ | Sample size. |
| $\bar{X}, \bar{Z}$ | Population means of auxiliary variables. |
| $\bar{Y}$ | Population mean of the study variable. |
| $\bar{x}, \bar{z}$ | Sample means of auxiliary variables. |
| $\bar{y}$ | Sample mean of the study variable. |
| $S_x^2, S_z^2$ | Population variances of X and Z, respectively. |
| $S_y^2$ | Population variance of Y. |
| $S_{XY}, S_{YZ}$ | Population covariance between X and Y, Z and Y, respectively. |
| $S_{XZ}$ | Population covariance between X and Z. |
| $\rho_{yx}, \rho_{yz}$ | Population correlation coefficients between X and Y, Z and Y, respectively. |
| $\rho_{xz}$ | Population correlation coefficients between X and Z. |
| $C_x, C_z$ | Coefficients of variation of $X$ and Z, respectively. |
| $C_y$ | Coefficient of variation of $Y$. |

In order to examine the bias and MSE properties of the suggested estimator relative to conventional estimators, the following error terms are introduced.

Let $e_0 = \frac{(\bar{y}-\bar{Y})}{\bar{Y}}$, $e_1 = \frac{(\bar{x}-\bar{X})}{\bar{X}}$, and $e_2 = \frac{(\bar{z}-\bar{Z})}{\bar{Z}}$, such that $E(e_i) = 0$ for $i = 0, 1, 2$.

$E(e_0^2) = \theta C_y^2$, $E(e_1^2) = \theta C_x^2$, and $E(e_2^2) = \theta C_z^2$.

$E(e_0 e_1) = \theta \rho_{yx} C_y C_x$, $E(e_0 e_2) = \theta \rho_{yz} C_y C_z$ and $E(e_1 e_2) = \theta \rho_{xz} C_x C_z$.

Where, $\theta = \left(\frac{1}{n} - \frac{1}{N}\right)$, $C_x^2 = \frac{S_x^2}{\bar{X}^2}$, $C_y^2 = \frac{S_y^2}{\bar{Y}^2}$ and $C_z^2 = \frac{S_z^2}{\bar{Z}^2}$.

$\rho_{yx} = \frac{S_{XY}}{S_X S_Y}$, $\rho_{yz} = \frac{S_{YZ}}{S_Y S_Z}$, and $\rho_{xz} = \frac{S_{XZ}}{S_X S_Z}$.

To assess the performance of the proposed PCA-based estimator, both existing and proposed estimators are considered for comparison.

## 3. Existing Estimator

This section provides a detailed description of certain established estimators for the population mean.

The usual estimator of the population mean and its corresponding variance are outlined below.

$$t_0 = \bar{y}. \tag{3.1}$$

$$V(t_0) = \theta \bar{Y}^2 C_y^2. \tag{3.2}$$

According to Singh and Chaudhary [21], the ratio-cum-ratio estimator, and its associated bias and MSE are given as follows.

$$t_1 = \bar{y}\left(\frac{\bar{X}}{\bar{x}}\right)\left(\frac{\bar{Z}}{\bar{z}}\right). \tag{3.3}$$

$$Bias(t_1) \cong \bar{Y}\theta\left(C_y^2 + C_z^2 - \rho_{yx}C_yC_x - \rho_{yz}C_yC_z + \rho_{xz}C_xC_z\right). \tag{3.4}$$

$$MSE(t_1) \cong \bar{Y}^2\theta\left(C_x^2 + C_y^2 + C_z^2 - 2\rho_{yx}C_yC_x - 2\rho_{yz}C_yC_z + 2\rho_{xz}C_xC_z\right). \tag{3.5}$$

Motivated by Singh[22], the ratio-cum-product estimator along with its corresponding bias and MSE is presented as follows.

$$t_2 = \bar{y}\left(\frac{\bar{X}}{\bar{x}}\right)\left(\frac{\bar{z}}{\bar{Z}}\right). \tag{3.6}$$

$$Bias(t_2) \cong \bar{Y}\theta\left(C_x^2 - \rho_{yx}C_yC_x - \rho_{yz}C_yC_z + \rho_{xz}C_xC_z\right). \tag{3.7}$$

$$MSE(t_2) \cong \bar{Y}^2\theta\left(C_x^2 + C_y^2 + C_z^2 - 2\rho_{yx}C_yC_x + 2\rho_{yz}C_yC_z - 2\rho_{xz}C_xC_z\right). \tag{3.8}$$

The product-cum-product estimator along with the corresponding expressions for bias and MSE is described as follows.

$$t_3 = \bar{y}\left(\frac{\bar{x}}{\bar{X}}\right)\left(\frac{\bar{z}}{\bar{Z}}\right). \tag{3.9}$$

$$Bias(t_3) \cong \bar{Y}\theta\left(\rho_{yx}C_yC_x + \rho_{yz}C_yC_z + \rho_{xz}C_xC_z\right). \tag{3.10}$$

$$MSE(t_3) \cong \bar{Y}^2\theta\left(C_x^2 + C_y^2 + C_z^2 + 2\rho_{yx}C_yC_x + 2\rho_{yz}C_yC_z + 2\rho_{xz}C_xC_z\right). \tag{3.11}$$

Inspired by Singh et al. [5], the formulation of the ratio-cum-product type exponential estimator along with its bias and MSE is presented below.

$$t_4 = \bar{y}\exp\left(\frac{\bar{X}-\bar{x}}{\bar{X}+\bar{x}}\right)\exp\left(\frac{\bar{z}-\bar{Z}}{\bar{Z}+\bar{z}}\right). \tag{3.12}$$

$$Bias(t_4) \cong \bar{Y}\theta\left(\frac{3}{8}C_x^2 - \frac{1}{4}C_z^2 - \frac{1}{4}\rho_{yx}C_yC_x + \frac{1}{2}\rho_{yz}C_yC_z - \frac{1}{4}\rho_{xz}C_xC_z\right). \tag{3.13}$$

$$MSE(t_4) \cong \bar{Y}^2\theta\left(\frac{1}{4}C_x^2 + C_y^2 + \frac{1}{4}C_z^2 - 2\rho_{yx}C_yC_x + 2\rho_{yz}C_yC_z - \frac{1}{2}\rho_{xz}C_xC_z\right). \tag{3.14}$$

The multivariate regression-type estimator proposed by Raj [14] along with its associated minimum MSE is presented as follows.

$$t_5 = \bar{y} + \beta_1(\bar{X} - \bar{x}) + \beta_2(\bar{Z} - \bar{z}). \tag{3.15}$$

Since $\beta_1$ and $\beta_2$ are unknown their optimal values are obtained as $\beta_1 = \frac{\bar{Y}C_y(\rho_{yx} - \rho_{xz}\rho_{yz})}{\bar{X}C_x(1-\rho_{xz}^2)}$ and $\beta_2 = \frac{\bar{Y}C_y(\rho_{yz} - \rho_{xz}\rho_{yx})}{\bar{Z}C_z(1-\rho_{xz}^2)}$ respectively.

$$MSE(t_5) = \bar{Y}^2\theta C_y^2\left(1 - R_{y.xz}^2\right). \tag{3.16}$$

where $R^2_{y.xz} = \frac{\rho^2_{yx}+\rho^2_{yz}-2\rho_{yx}\rho_{yz}\rho_{xz}}{1-\rho^2_{xz}}$ refers to the multiple correlation between the variables $Y$, $X$ and $Z$.

## 4. Proposed Estimator

### 4.1 Proposed Logarithmic-Type Estimator

This section presents an improved estimator for the finite population mean under the simple random sampling framework. Motivated by the works of Nasir and Ahmad [23], the proposed estimator incorporates suitable functions of known parameters of auxiliary variables to enhance estimation efficiency. The proposed log-type estimator is defined as follows:

$$t^* = \bar{y} + K\log\left[\frac{\bar{X}+\psi_1}{\bar{x}+\psi_1}\right] + (1-K)\log\left[\frac{\bar{z}+\psi_2}{\bar{Z}+\psi_2}\right] \quad (4.1)$$

where $\psi_1$ and $\psi_2$ are either real constants or functions of known parameters of the auxiliary variables $X$ and $Z$, respectively. Such parameters may include the coefficients of kurtosis $\beta_2(x)$ and $\beta_2(z)$, the population correlation coefficients $\rho_{yx}$ and $\rho_{yz}$, the standard deviations $S_x$ and $S_z$, and the coefficients of variation $C_x$ and $C_z$. The constant $K$ is evaluated by minimizing the MSE associated with the estimator $t^*$.

The first-order expansion of equation (4.1) in terms of the error components ($e's$) gives the following result:

$$t^* = (1+e_0)\bar{Y} + K\log(1+e_1\eta_1)^{-1} + (1-K)\log(1+e_2\eta_2). \quad (4.2)$$

Where, $\eta_1 = \frac{\bar{X}}{\bar{X}+\psi_1}$ and $\eta_2 = \frac{\bar{Z}}{\bar{Z}+\psi_2}$.

Following expansion and simplification of the right-hand side, the result is given by:

$$t^* = (1+e_0)\bar{Y} + K\log(1-e_1\eta_1+e_1^2\eta_1^2) + (1-K)\log(1+e_2\eta_2). \quad (4.3)$$

$$t^* = (1+e_0)\bar{Y} + K\left(\frac{e_1^2\eta_1^2}{2} - e_1\eta_1\right) + (1-K)\left(e_2\eta_2 - \frac{(e_2\eta_2)^2}{2}\right). \quad (4.4)$$

After subtracting $\bar{Y}$ from both sides of equation (4.4), the following expression is obtained:

$$t^* - \bar{Y} = \bar{Y}e_0 + K\left(\frac{e_1^2\eta_1^2}{2} - e_1\eta_1\right) + (1-K)\left(e_2\eta_2 - \frac{(e_2\eta_2)^2}{2}\right). \quad (4.5)$$

Taking expectations on both sides of equation (4.5) and substituting the corresponding values yields the bias of estimator $t^*$ as:

$$E(t^* - \bar{Y}) = \bar{Y}E(e_0) + K\left(\frac{E(e_1^2)\eta_1^2}{2} - E(e_1)\eta_1\right) + (1-K)\left(E(e_2)\eta_2 - \frac{E(e_2^2)\eta_2^2}{2}\right). \quad (4.6)$$

$$Bias(t^*) = K\left(\frac{\eta_1^2}{2}\theta C_x^2\right) - (1-K)\left(\frac{\eta_2^2}{2}\theta C_z^2\right). \quad (4.7)$$

By utilizing equation (4.5), we derive:

$$t^* - \bar{Y} \cong \bar{Y}e_0 - Ke_1\eta_1 + (1-K)\eta_2 e_2. \quad (4.8)$$

The MSE of estimator $t^*$ to the first-order approximation is obtained by squaring equation (4.8) and evaluating its expectation:

$$MSE(t^*) = \begin{bmatrix} (\bar{Y}^2\theta C_y^2 + \eta_2^2\theta C_z^2 + 2\bar{Y}\eta_2\theta\rho_{yz}C_yC_z) + K^2\begin{pmatrix} \eta_1^2\theta C_x^2 + \eta_2^2\theta C_z^2 + \\ 2\eta_1\eta_2\theta\rho_{xz}C_xC_z \end{pmatrix} - \\ 2K(\eta_2^2\theta C_z^2 + \bar{Y}\eta_1\theta\rho_{yx}C_yC_x + \bar{Y}\eta_2\theta\rho_{yz}C_yC_z + \eta_1\eta_2\theta\rho_{xz}C_xC_z) \end{bmatrix} \quad (4.9)$$

$$MSE(t^*) = A + BK^2 - 2KC. \quad (4.10)$$

Where,

$A = \bar{Y}^2\theta C_y^2 + \eta_2^2\theta C_z^2 + 2\bar{Y}\eta_2\theta\rho_{yz}C_yC_z.$

$B = \eta_1^2\theta C_x^2 + \eta_2^2\theta C_z^2 + 2\eta_1\eta_2\theta\rho_{xz}C_xC_z.$

$C = \eta_2^2\theta C_z^2 + \bar{Y}\eta_1\theta\rho_{yx}C_yC_x + \bar{Y}\eta_2\theta\rho_{yz}C_yC_z + \eta_1\eta_2\theta\rho_{xz}C_xC_z.$

The optimal value of $K$ that minimizes the MSE of estimator $t^*$ is given by:

$$K_{opt} = \frac{C}{B}.$$

Substituting the optimal value $K_{opt}$, the minimum MSE of the estimator $t^*$ is given by:

$$MSE_{min}(t^*) = \left[A - \frac{C^2}{B}\right]. \quad (4.11)$$

### 4.2 Proposed PCA-Based Estimator under Multicollinearity

Principal Component Analysis (PCA) provides a reliable approach to handle the aforementioned issues. PCA is a statistical technique that transforms a set of correlated auxiliary variables into a new set of linearly uncorrelated variables, referred to as principal components. These components are ranked based on their contribution to the total variance in the data. By retaining only the most significant components, PCA reduces dimensionality, eliminates redundancy, and mitigates multicollinearity, leading to more stable and efficient estimation of the population mean. To formalize the proposed estimation framework, the notation below is employed.

Let $Y$ denote the variable of interest with finite population mean $\bar{Y}$. Additionally, let $X = [X_1, X_2, \ldots, X_p]$ denote the matrix comprising $p$ auxiliary variables for the population. The matrix $Z = [Z_1, Z_2, \ldots, Z_k]$ consists of the transformed principal components obtained from $X$, where $k \leq p$ corresponds to the number of components preserved after dimensionality reduction.

In the present study, we consider $p = 2$ auxiliary variables and retain only the first principal component ($k = 1$), as it captures the majority of the variability in the auxiliary information. The construction of the PCA-based ratio estimator begins by standardizing the auxiliary variables to have zero mean and unit variance, especially when they are measured on different scales. After standardizing the variables, the covariance matrix is evaluated to capture the linear associations among the auxiliary variables. It is defined as:

$$\Sigma_x = \frac{1}{N} X^T X.$$

where $X^T$ represents the transpose of the standardized matrix $X$, and $N$ denotes the total number of observations.

The subsequent step involves performing eigen decomposition of the covariance matrix, expressed as $\Sigma_x = P \Delta P^T$,

where $P$ denotes the matrix of eigenvectors defining the directions of the orthogonal principal components, and $\Delta = diag(\lambda_1, \lambda_2, \ldots, \lambda_P)$ is the diagonal matrix of eigenvalues arranged such that $\lambda_1 \geq \lambda_2 \geq \ldots \geq \lambda_P$.

The reduction in dimensionality is carried out by selecting the first $k$ principal components that account for a considerable fraction of the total variance, computed as

$$W = X P_k.$$

where $P_k$ consists of the eigenvectors corresponding to the $k$ largest eigenvalues. In this study, since $k = 1$, the transformation reduces to a single principal component. The resulting component $W$ denotes the matrix of principal components. These uncorrelated variables are subsequently used in the suggested estimator to mitigate multicollinearity and improve stability.

In this study, only the first principal component is retained ($k = 1$) from the two auxiliary variables, as it captures the majority of the variability in the auxiliary information. Based on this component, we propose a log-type estimator that incorporates the transformed uncorrelated variable to improve stability and mitigate multicollinearity. The proposed estimator is expressed as:

$$t_{PCA} = \bar{y} + \alpha \log\left[\frac{\overline{W}+\psi}{\bar{w}+\psi}\right]. \qquad (4.12)$$

Where $\psi$ is a real constant or a function of known parameters of the principal component $W$, such as the coefficient of kurtosis $\beta_2(w)$, the population correlation coefficient $\rho_{yw}$, the standard deviation $S_w$, and the coefficient of variation $C_w$. The constant $\alpha$ is chosen such that the MSE of the estimator $t_{PCA}$ is minimized.

To examine the bias and MSE properties of the proposed PCA-based estimator, the relative error components are defined as follows:

Let $e_0 = \frac{(\bar{y}-\bar{Y})}{\bar{Y}}$, $e_w = \frac{(\bar{w}-\overline{W})}{\overline{W}}$, such that $E(e_0) = E(e_w) = 0$.

$E(e_0^2) = \theta C_y^2, E(e_w^2) = \theta C_w^2$, and $E(e_0 e_w) = \theta \rho_{yw} C_y C_w$.

By expressing the proposed estimator given in equation (4.12) in terms of the above relative error components and applying first-order approximation, the following expression is obtained.

$$t_{PCA} = (1+e_0)\bar{Y} + \alpha \log(1 + e_w \eta_3)^{-1}. \qquad (4.13)$$

Where, $\eta_3 = \frac{\overline{W}}{\overline{W}+\psi}$.

After simplifying the expanded form of the right-hand side, we get:

$$t_{PCA} = (1+e_0)\bar{Y} + \alpha \log(1 - e_w \eta_3 + e_w^2 \eta_3^2). \qquad (4.14)$$

$$t_{PCA} = (1+e_0)\bar{Y} + \alpha \left(\frac{e_w^2 \eta_3^2}{2} - e_w \eta_3\right). \qquad (4.15)$$

Subtracting $\bar{Y}$ from both sides of equation (4.15) yields:

$$t_{PCA} - \bar{Y} = \bar{Y}e_0 + \alpha \left(\frac{e_w^2 \eta_3^2}{2} - e_w \eta_3\right). \qquad (4.16)$$

Taking expectations on both sides of equation (4.16) and substituting the corresponding values yields the bias of estimator $t_{PCA}$ as:

$$Bias(t_{PCA}) = \alpha \left(\frac{\eta_3^2}{2} \theta C_w^2\right). \qquad (4.17)$$

Equation (4.16) yields the following expression:

$$t_{PCA} - \bar{Y} \cong \bar{Y}e_0 - \alpha e_w \eta_3. \qquad (4.18)$$

Squaring both sides of equation (4.18) and taking expectations yields the first-order MSE of the estimator $t_{PCA}$ as:

$$MSE(t_{PCA}) = \bar{Y}^2 \theta C_y^2 + \alpha^2 \eta_3^2 \theta C_w^2 - 2\alpha \bar{Y} \eta_3 \theta \rho_{yw} C_y C_w. \qquad (4.19)$$

$$MSE(t_{PCA}) = A_1 + B_1 \alpha^2 - 2\alpha C_1. \qquad (4.20)$$

Where,

$A_1 = \bar{Y}^2 \theta C_y^2.$

$B_1 = \eta_3^2 \theta C_w^2.$

$C_1 = \bar{Y} \eta_3 \theta \rho_{yw} C_y C_w.$

The MSE of estimator $t_{PCA}$ is minimized at the following optimal value of $\alpha$:

$$\alpha_{opt} = \frac{C_1}{B_1}.$$

Substituting the optimal value $\alpha_{opt}$, the minimum MSE of the estimator $t_{PCA}$ is given by:

$$MSE_{min}(t_{PCA}) = \left[A_1 - \frac{C_1^2}{B_1}\right]. \qquad (4.21)$$

The overall procedure of the proposed PCA-based estimator under multicollinearity is illustrated in Figure 1.

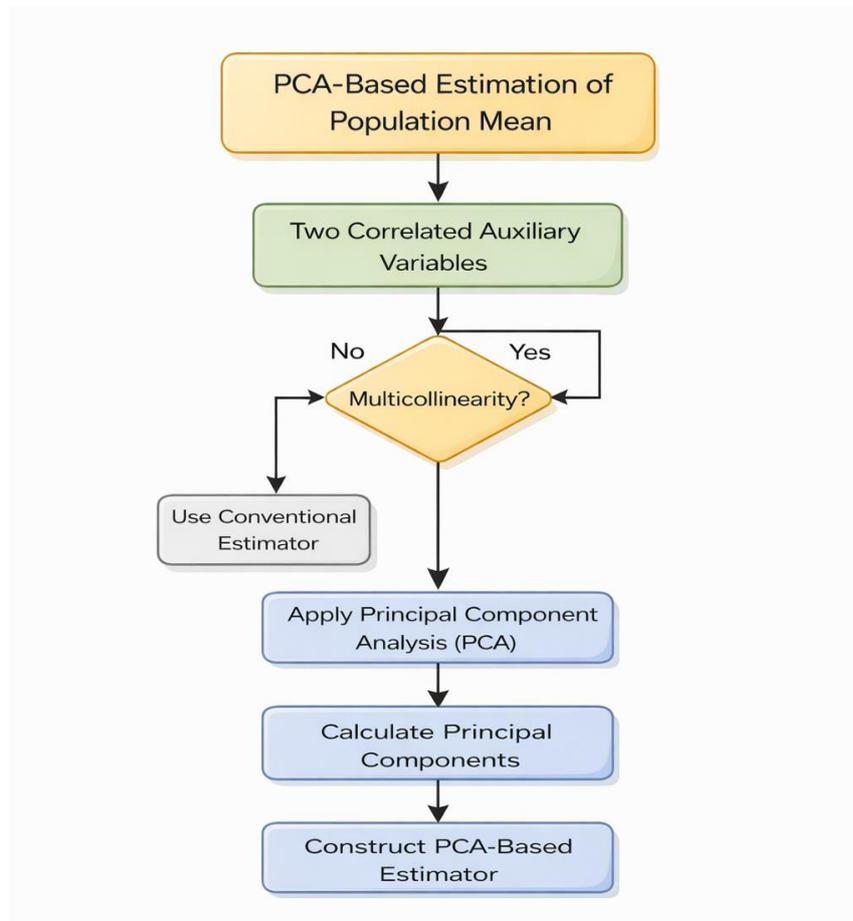

**Figure 1:** Flowchart of the proposed PCA-based estimation procedure.

The flowchart presents the sequential steps involved in constructing the proposed estimator and addressing multicollinearity through the application of PCA.

## 5. Quantitative Assessment

In this section, the performance of the proposed estimator $(t^*)$ and the PCA-based estimator $(t_{PCA})$ is examined using real dataset. The efficiency of these estimators is compared with conventional estimators based on MSE and PRE. The objective is to assess the effectiveness of the proposed approach in practical situations, particularly in the presence of multicollinearity among auxiliary variables. A detailed description of the datasets along with their key statistical characteristics is provided below.

**Dataset**: (Source: Maddala and Lahiri[24], p.286).

$Y$:Imports (in millions of New Francs at 1959 prices).

$X$:Gross Domestic Product (in millions of New Francs at 1959 prices).

$Z$: Consumption (in millions of New Francs at 1959 prices).

The descriptive statistical characteristics of the dataset are outlined as follows.

$N = 18, n = 10, \bar{Y} = 30.078, \bar{X} = 237.517, \bar{Z} = 167.378, \bar{W} = 0.0015, C_y = 0.4033, C_x = 0.2599, C_z = 0.2414, C_W = 914.8534, \rho_{yx} = 0.9842, \rho_{yz} = 0.9848, \rho_{yW} = 0.9847.$

**Table 2:** Multicollinearity diagnostics of auxiliary variables.

| $\rho_{xz}$ | VIF | CI | eig1 | eig2 |
|---|---|---|---|---|
| 0.9989 | 468.8343 | 43.28206 | 1.9989 | 0.0012 |

Table 2 presents the values of correlation coefficient, VIF, condition index (CI), and eigenvalues. The high correlation and large VIF indicate the presence of severe multicollinearity among the auxiliary variables.

The extent of multicollinearity can be quantified using the variance inflation factor (VIF), which is defined as

$$VIF = \frac{1}{1-\rho_{xz}^2}.$$

where $\rho_{xz}$ denotes the correlation coefficient between the auxiliary variables $X$ and $Z$. The VIF measures the degree to which the variance of an estimator is inflated due to multicollinearity. A larger value of VIF indicates a higher level of multicollinearity and consequently greater instability in the estimation process.

**Table 3:** Comparative analysis of estimators using MSE and PRE for the real dataset.

| Estimator | MSE | PRE |
|---|---|---|
| $t_0$ | 6.5390 | 100.00 |
| $t_1$ | 0.6337 | 1031.89 |
| $t_2$ | 5.9735 | 109.47 |
| $t_3$ | 32.6432 | 20.03 |
| $t_4$ | 5.9593 | 109.73 |
| $t_5$ | 0.3882 | 1684.43 |
| $t^*$ | 0.1984 | 3296.60 |
| **$t_{PCA}$** | **0.1974** | **3310.96** |

Percent relative efficiency (PRE) evaluates the relative performance of the proposed estimators with respect to the usual mean estimator based on their MSE values. A higher PRE indicates greater efficiency of the proposed estimators in reducing MSE. The PRE is computed using the following formula:

$$PRE(Estimator) = \frac{MSE(t_0)}{MSE(Estimator)} \times 100.$$

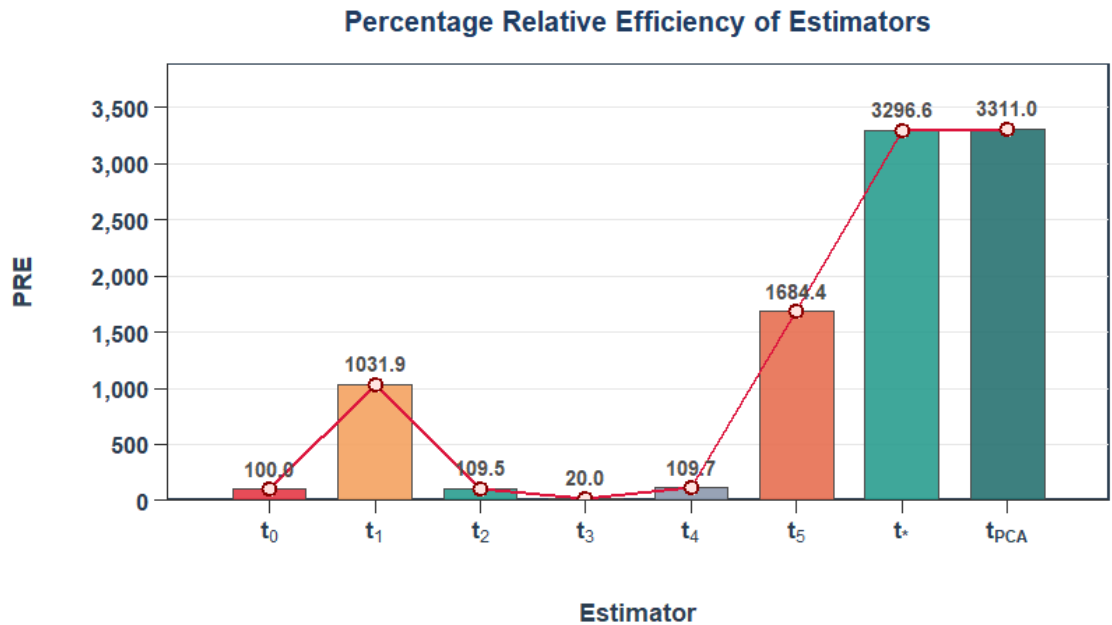

**Figure 2**: PRE outcomes of the estimators illustrated using real dataset.

The line graph depicting the PRE values of various estimators for the real dataset are presented in Figure 2. These graphical representations facilitate a clear comparison of estimator performance, highlighting the superior efficiency of the proposed estimator ($t^*$) and the PCA-based estimator ($t_{PCA}$) over conventional estimators.

## 6. Simulation Study

A comprehensive simulation study is conducted to evaluate the performance of the proposed estimator in the presence of multicollinearity among auxiliary variables. Several existing estimators are also considered for comparison. Furthermore, a Principal Component Analysis (PCA)-based estimator is incorporated to examine its effectiveness in mitigating the adverse effects of multicollinearity. The simulation is carried out under a finite population framework, where correlated auxiliary variables are generated with varying levels of correlation.

The performance of the estimators is assessed using mean squared error (MSE) and percent relative efficiency (PRE). The following steps are adopted in the simulation process:

a) We employed the R programming language to simulate a tri-variate normal distribution for $Y, X$, and $Z$ with a population size of $N = 1000$ based on the specified parameters.

b) The tri-variate normal distribution is generated using a specified mean vector and covariance structure. The mean vector is defined as $[\mu_x, \mu_y, \mu_z] = [12, 15, 19]$, while the covariance matrix is given by:

$$\Sigma = \begin{bmatrix} \sigma_y^2 & \rho_{xy}\sigma_y\sigma_x & \rho_{yz}\sigma_y\sigma_z \\ \rho_{xy}\sigma_y\sigma_x & \sigma_x^2 & \rho_{xz}\sigma_x\sigma_z \\ \rho_{yz}\sigma_y\sigma_z & \rho_{xz}\sigma_x\sigma_z & \sigma_z^2 \end{bmatrix}$$

The variances are taken as $\sigma_x^2 = 36$, $\sigma_y^2 = 16$, and $\sigma_z^2 = 9$. The correlation coefficients are fixed at $\rho_{yx} = 0.75$ and $\rho_{yz} = 0.76$, while $\rho_{xz}$ is varied from 0.3 to 0.9 to represent different levels of correlation among the auxiliary variables.

c) To address multicollinearity between the auxiliary variables $X$ and $Z$, principal component analysis (PCA) is applied to obtain an uncorrelated linear combination. Specifically, the first principal component is constructed as:

$$W = P_1 X + P_2 Z$$

where $P_1$ and $P_2$ are the normalized eigenvector coefficients corresponding to the largest eigenvalue of the covariance matrix of $X$ and $Z$.

d) All relevant statistical parameters were derived from the simulated population.

e) Samples of sizes n = (50, 100, 150) were drawn from the simulated population using SRSWOR.

f) The performance of the proposed and existing estimators was evaluated in terms of MSE and PRE for varying values of $\rho_{xz}$. Furthermore, variance inflation factor (VIF), the condition index (CI), and eigenvalues were computed to quantify the level of multicollinearity.

g) The simulation was replicated 25,000 times, and the MSE for each estimator of the population mean was obtained by averaging the 25,000 computed values.

**Table 4:** Multicollinearity diagnostics of auxiliary variables for $n = 50$.

| $\rho_{xz}$ | VIF | CI | eig1 | eig2 |
|---|---|---|---|---|
| 0.3 | 1.0989 | 1.2699 | 1.2345 | 0.7655 |
| 0.4 | 1.1905 | 1.5521 | 1.4133 | 0.5867 |
| 0.5 | 1.3333 | 1.7642 | 1.5137 | 0.4863 |
| 0.6 | 1.5625 | 1.9555 | 1.5854 | 0.4146 |
| 0.7 | 1.9608 | 2.4840 | 1.7211 | 0.2789 |
| 0.8 | 2.7778 | 3.0925 | 1.8107 | 0.1893 |
| 0.9 | 5.2632 | 4.2497 | 1.8951 | 0.1049 |

Table 4 demonstrates a clear pattern in which higher correlation $\rho_{xz}$ results in increased VIF and CI values and a decrease in the smallest eigenvalue, indicating stronger multicollinearity.

The condition index (CI), defined as the square root of the ratio of the largest to the smallest eigenvalue of the correlation matrix, is used to assess the numerical stability of the system. A higher value of CI indicates the presence of multicollinearity and potential instability in estimation.

$$CI = \sqrt{\frac{\lambda_{max}}{\lambda_{min}}}.$$

where $\lambda_{max}$ and $\lambda_{min}$ denote the largest and smallest eigenvalues, respectively.

**Table 5:** Mean squared error (MSE) of estimators under multicollinearity for $n = 50$.

| $\rho_{xz}$ | $t_0$ | $t_1$ | $t_2$ | $t_3$ | $t_4$ | $t_5$ | $t^*$ | $t_{PCA}$ |
|---|---|---|---|---|---|---|---|---|
| 0.3 | 0.2983 | 0.1900 | 0.4103 | 1.7258 | 0.2229 | 0.0427 | 0.0791 | 0.0426 |
| 0.4 | 0.2940 | 0.2122 | 0.3445 | 1.6978 | 0.2209 | 0.0604 | 0.0863 | 0.0625 |
| 0.5 | 0.3134 | 0.2226 | 0.3389 | 1.7630 | 0.2440 | 0.0804 | 0.1036 | 0.0792 |
| 0.6 | 0.2946 | 0.2964 | 0.2751 | 1.8066 | 0.2033 | 0.1065 | 0.1152 | 0.1050 |
| 0.7 | 0.2886 | 0.2923 | 0.2120 | 1.7817 | 0.1885 | 0.0996 | 0.1023 | 0.0992 |
| 0.8 | 0.3019 | 0.2992 | 0.1985 | 1.7676 | 0.1994 | 0.1168 | 0.1182 | 0.1146 |
| 0.9 | 0.2887 | 0.2918 | 0.1579 | 1.7102 | 0.1785 | 0.1207 | 0.1207 | 0.1194 |

The results indicate that the proposed PCA-based estimator consistently yield lower MSE values compared to the existing estimators across different levels of $\rho_{xz}$, demonstrating their superior performance in the presence of multicollinearity.

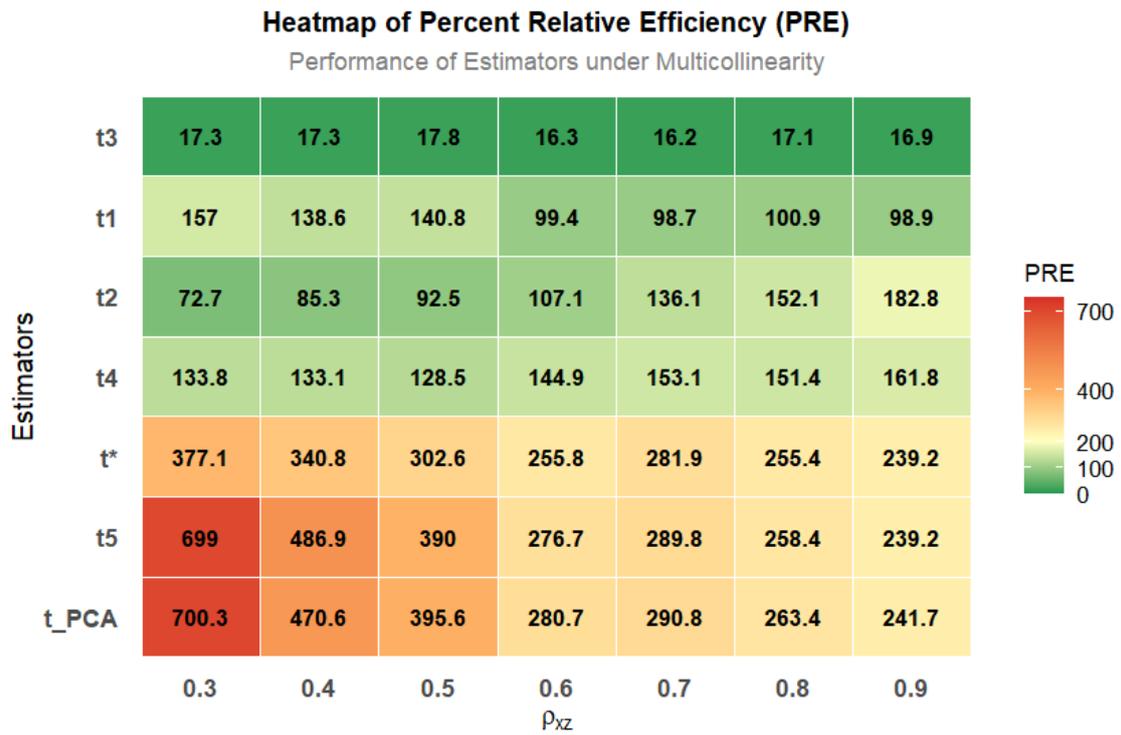

**Figure 3:** Heatmap of percent relative efficiency (PRE) of existing and proposed estimators under multicollinearity for $n = 50$.

**Table 6:** Multicollinearity diagnostics of auxiliary variables for $n = 100$.

| $\rho_{xz}$ | VIF | CI | eig1 | eig2 |
|---|---|---|---|---|
| 0.3 | 1.0989 | 1.2699 | 1.2345 | 0.7655 |
| 0.4 | 1.1905 | 1.5571 | 1.4160 | 0.5840 |
| 0.5 | 1.3333 | 1.8019 | 1.5291 | 0.4709 |
| 0.6 | 1.5625 | 2.0514 | 1.6160 | 0.3840 |
| 0.7 | 1.9608 | 2.2672 | 1.6743 | 0.3257 |
| 0.8 | 2.7778 | 3.1717 | 1.8192 | 0.1808 |
| 0.9 | 5.2632 | 4.3170 | 1.8981 | 0.1019 |

It is evident from Table 6 that increasing values of $\rho_{xz}$ lead to higher VIF and CI values, along with a decline in the smaller eigenvalue, indicating a growing severity of multicollinearity among the auxiliary variables.

**Table 7:** Mean squared error (MSE) of estimators under multicollinearity for $n = 100$.

| $\rho_{xz}$ | $t_0$ | $t_1$ | $t_2$ | $t_3$ | $t_4$ | $t_5$ | $t^*$ | $t_{PCA}$ |
|---|---|---|---|---|---|---|---|---|
| 0.3 | 0.1304 | 0.0772 | 0.1785 | 0.7445 | 0.1007 | 0.0189 | 0.0347 | 0.0189 |

| 0.4 | 0.1487 | 0.0875 | 0.1446 | 0.8474 | 0.1048 | 0.0252 | 0.0357 | 0.0258 |
| 0.5 | 0.1471 | 0.0970 | 0.1238 | 0.8370 | 0.0967 | 0.0366 | 0.0414 | 0.0364 |
| 0.6 | 0.1369 | 0.1226 | 0.1258 | 0.8294 | 0.0955 | 0.0419 | 0.0462 | 0.0410 |
| 0.7 | 0.1383 | 0.1432 | 0.1206 | 0.8104 | 0.0993 | 0.0550 | 0.0591 | 0.0551 |
| 0.8 | 0.1327 | 0.1539 | 0.0765 | 0.8919 | 0.0779 | 0.0464 | 0.0469 | 0.0460 |
| 0.9 | 0.1490 | 0.1680 | 0.0737 | 0.9873 | 0.0863 | 0.0548 | 0.0546 | 0.0536 |

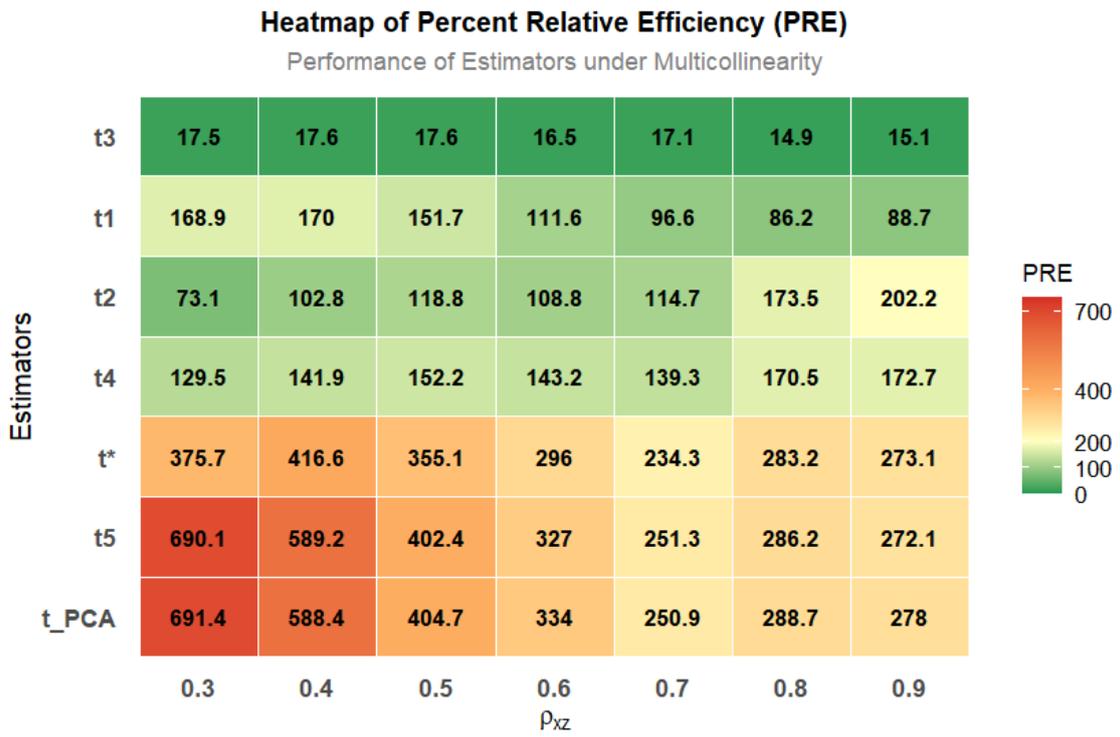

**Figure 4:** Heatmap of percent relative efficiency (PRE) of existing and proposed estimators under multicollinearity for $n = 100$.

**Table 8:** Multicollinearity diagnostics of auxiliary variables for $n = 150$.

| $\rho_{xz}$ | VIF | CI | eig1 | eig2 |
| --- | --- | --- | --- | --- |
| 0.3 | 1.0989 | 1.2699 | 1.2345 | 0.7655 |
| 0.4 | 1.1905 | 1.5373 | 1.4053 | 0.5947 |
| 0.5 | 1.3333 | 1.7979 | 1.5275 | 0.4725 |
| 0.6 | 1.5625 | 1.8605 | 1.5517 | 0.4483 |
| 0.7 | 1.9608 | 2.4454 | 1.7135 | 0.2865 |
| 0.8 | 2.7778 | 2.9651 | 1.7957 | 0.2043 |
| 0.9 | 5.2632 | 4.3618 | 1.9001 | 0.0999 |

**Table 9:** Mean squared error (MSE) of estimators under multicollinearity for $n = 150$.

| $\rho_{xz}$ | $t_0$ | $t_1$ | $t_2$ | $t_3$ | $t_4$ | $t_5$ | $t^*$ | $t_{PCA}$ |
|---|---|---|---|---|---|---|---|---|
| 0.3 | 0.0749 | 0.0491 | 0.1109 | 0.4364 | 0.0591 | 0.0122 | 0.0214 | 0.0123 |
| 0.4 | 0.1006 | 0.0649 | 0.1053 | 0.6016 | 0.0705 | 0.0181 | 0.0249 | 0.0184 |
| 0.5 | 0.1026 | 0.0631 | 0.0927 | 0.5707 | 0.0713 | 0.0229 | 0.0291 | 0.0230 |
| 0.6 | 0.0883 | 0.0817 | 0.0883 | 0.5285 | 0.0654 | 0.0293 | 0.0331 | 0.0289 |
| 0.7 | 0.0861 | 0.0861 | 0.0691 | 0.5324 | 0.0578 | 0.0294 | 0.0318 | 0.0293 |
| 0.8 | 0.0882 | 0.0989 | 0.0660 | 0.5545 | 0.0607 | 0.0317 | 0.0344 | 0.0313 |
| 0.9 | 0.0901 | 0.1134 | 0.0482 | 0.6125 | 0.0548 | 0.0339 | 0.0346 | 0.0335 |

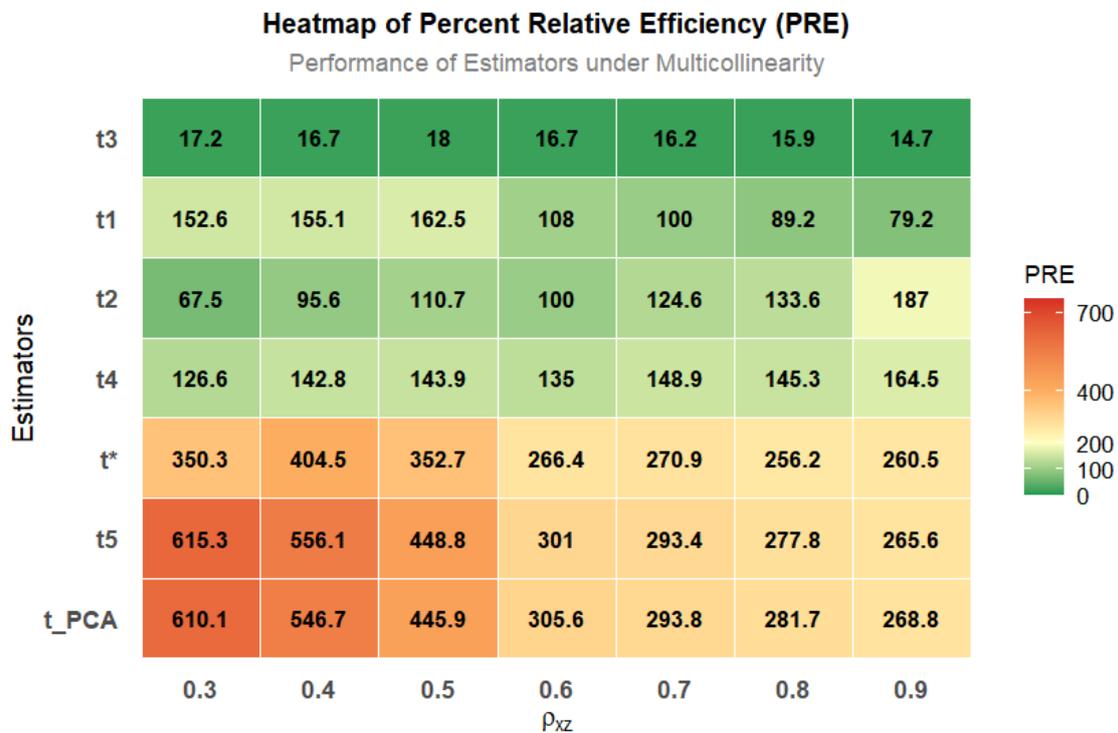

**Figure 5:** Heatmap of percent relative efficiency (PRE) of existing and proposed estimators under multicollinearity for $n = 150$.

The findings presented in Tables 5, 7, and 9 demonstrate the performance of the estimators based on their MSE values.

## 7. Results and Discussion

The results presented in Table 2 indicate a very high correlation ($\rho_{xz} = 0.9989$) between the auxiliary variables, suggesting the presence of severe multicollinearity. This is further supported by the large value of the variance inflation factor (VIF = 468.8343) and a high condition index (CI = 43.28206). Additionally, the eigenvalues show a substantial disparity, confirming the existence of near-linear dependence among the variables.The results in Table 3 indicate a substantial improvement in estimation efficiency with the use of auxiliary information. The conventional estimator $t_0$ shows the highest MSE (6.5390) and serves as the baseline, while estimators $t_1$ and $t_5$ exhibit notable efficiency gains. In contrast, $t_3$ performs poorly due to its product-type structure under positive correlation, resulting in a large MSE.The proposed estimator $t^*$ achieves a significantly lower MSE (0.1984) and high PRE (3296.60), while the PCA-based estimator $t_{PCA}$ attains the minimum MSE (0.1974) and highest PRE (3310.96), reflecting superior performance. Overall, these results confirm that the proposed and PCA-based estimators outperform existing methods, with the PCA approach providing the most efficient estimates.

A simulation study based on the tri-variate normal distribution was conducted to validate the empirical findings.Tables 4, 6, and 8 summarize the values of the VIF, CI, and eigenvalues of the correlation matrix across different levels of correlation between auxiliary variables.As shown in Table 4, as the correlation coefficient $\rho_{xz}$ increases from 0.3 to 0.9, the VIF rises from 1.0989 to 5.2632 and the CI increases from 1.2699 to 4.2497, indicating growing multicollinearity and reduced numerical stability. The eigenvalues further support this pattern, with the largest increasing and the smallest decreasing toward zero, suggesting near-linear dependence among auxiliary variables. This reflects the presence of redundancy in the information conveyed. Similar trends observed in Tables 6 and 8 further substantiate these findings.

Overall, the combined behavior of VIF, CI, and eigenvalues indicates that increasing correlation among auxiliary variables intensifies multicollinearity, reduces numerical stability, and increases redundancy, thereby justifying the use of dimensionality reduction methods such as principal component analysis (PCA).Furthermore, the estimators were compared using MSE and PRE, with PRE derived from the MSE of the usual estimator. The findings show that the PCA-based estimator demonstrates superior performance across different levels of correlation among auxiliary variables. This is supported by MSE results in Tables 5, 7, and 9 and PRE visualizations in Figures 3–5, highlighting its robustness under multicollinearity.

## 8. Conclusion

This study proposes a PCA-based estimator for estimating the finite population mean under simple random sampling without replacement and compares its efficiency with the classical estimators in the presence of multicollinearity among auxiliary variables.The bias and MSE of the suggested estimator are obtained using first-order approximation methods.Both analytical results and numerical evaluation using a real dataset and a tri-variate normal simulation establish the improved efficiency of the proposed estimator over existing estimators in this study.The simulation results, based on theoretical MSE evaluated under varying levels of multicollinearity, consistently show the advantage of the PCA-based estimator, as reported in Tables 5, 7, and 9. These findings clearly indicate that the PCA-based estimator effectively mitigates the adverse effects of multicollinearity by transforming correlated auxiliary variables into orthogonal components, therebyreducing redundancy, and improving the stability and efficiency of population mean estimation.

## 9. Future Works

The present study can be extended in several directions to enhance the applicability of the proposed PCA-based estimator:

(1) Extending the estimator to incorporate multiple auxiliary variables in high-dimensional settings.
(2) Generalizing the methodology for estimating other population parameters such as variance, proportion, median, and distribution functions.
(3) Adapting the proposed approach to alternative sampling designs including stratified, systematic, cluster, and two-phase sampling.
(4) Examining the robustness of the estimator in the presence of measurement errors, missing data, and non-response.
(5) Integrating PCA with other dimension reduction or regularization techniques to further improve estimation efficiency.

**Data Availability**

All the data used for this study can be found inside the manuscript.

**Ethical Approval**

This research does not involve experimentation on Human Participants or Animals.

**Funding**

The authors declare that no funds or other grants were received for the preparation of this manuscript.

**Conflicts of Interest**

The authors declare no conflicts of interest.

**References**

[1] Bahl, S., & Tuteja, R. K. (1991). Ratio and product type exponential estimators. Journal of Information and Optimization Sciences, 12(1), 159–164. https://doi.org/10.1080/02522667.1991.10699058.

[2] Isaki, C. T., & Fuller, W. A. (1982). Survey design under the regression superpopulation model. Journal of the American Statistical Association, 77(377), 89–96. https://doi.org/10.1080/01621459.1982.10477770.

[3] Särndal, C. E., Swensson, B., & Wretman, J. (2003). Model assisted survey sampling. Springer Science & Business Media.

[4]   Kadilar, C., & Cingi, H. (2004). Ratio estimators in simple random sampling. Applied Mathematics and Computation, 151(3), 893–902. https://doi.org/10.1016/S0096-3003(03)00803-8.

[5]   Singh, H. P., Upadhyaya, L. N., & Tailor, R. (2009). Ratio-cum-product type exponential estimator. Statistica, 69(4), 299–310.

[6]   Sajjad, M., & Ismail, M. (2024). Efficient generalized estimators of population mean in the presence of non-response and measurement error. Kuwait Journal of Science, 51(3), 100224.

[7]   Shabbir, J., & Gupta, S. (2010). On estimating finite population mean in simple and stratified random sampling. Commun. Stat. - Theory Methods, 40(2), 199–212. https://doi.org/10.1080/03610920903411259.

[8]   Singh, R., Chauhan, P., Sawan, N., & Smarandache, F. (2007). Improvement in estimating the population mean using exponential estimator in simple random sampling. In Auxiliary Information and a Priori Values in Construction of Improved Estimators (Vol. 33).

[9]   Upadhyaya, L. N., & Singh, H. P. (1999). Use of transformed auxiliary variable in estimating the finite population mean. Biometrical Journal: Journal of Mathematical Methods in Biosciences, 41(5), 627-636.

[10]  Singh, R., Kumari, A., Smarandache, F., & Tiwari, S. N. (2025). Construction of almost unbiased estimator for population mean using neutrosophic information. Neutrosophic Sets and Systems, 76, 449–463. https: //doi.org/10.5281/zenodo.14010268.

[11]  Singh, R., & Tiwari, S. N. (2025). Improved estimator for population mean utilizing known medians of two auxiliary variables under neutrosophic framework. Neutrosophic systems with applications, 25(1), 38–52.

[12]  Singh, R., Kumari, A., Dubey, S., & Tiwari, S. N. (2025). Some novel sine-type estimators for finite population mean utilizing known auxiliary information. Quality & Quantity. https://doi.org/10.1007/s11135-025-02347-9.

[13]  Singh, P., Singh, A., & Sharma, P. (2025). A new class of logarithmic estimators using subsidiary information: Real-world applications and simulation insights. Sankhya B, 1–27.

[14]  Raj, D. (1965). On a method of using multi-auxiliary information in sample surveys. Journal of the American Statistical Association, 60(309), 270–277. https://doi.org/10.1080/01621459.1965.10480789.

[15]  Ahmad, Z., Hanif, M., & Maqsood, I. (2013). Generalized estimator of population mean for two-phase sampling using multi-auxiliary variables in the presence of non-response at first phase for no information case. Pakistan Journal of Statistics, 29(2).

[16] Sher, K., Ameeq, M., Hassan, M. M., Albalawi, O., & Afzal, A. (2024). Development of improved estimators of finite population mean in simple random sampling with dual auxiliaries and its application to real world problems. Heliyon, 10(10).

[17] Almulhim, F. A., et al. (2024). Estimation of finite population mean using dual auxiliary information under non-response with simple random sampling. Alexandria Engineering Journal, 100, 286–299.

[18] Kmenta, J., & Klein, L. R. (1971). Elements of econometrics (Vol. 655). New York: Macmillan.

[19] Gujarati, D. N. (2012). Basic Econometrics 4th ed.

[20] Jolliffe, I. (2025). Principal component analysis. In M. Lovric (Ed.), International encyclopedia of statistical science (pp. 1945–1948). Springer. doi: 10.1007/978-3-662-69359-9_483.

[21] Singh, D., & Chaudhary, F. S. (1986). Theory and analysis of sample survey designs. (No Title).

[22] Singh, M. P. (1967). Ratio cum product method of estimation. Metrika, 12(1), 34–42. https://doi.org/10.1007/BF02613481.

[23] Nasir, U., & Ahmad, Z. (2025). Enhancing population mean estimation through PCA-based estimators: Addressing multicollinearity and dimensionality challenges in survey sampling. Pakistan Journal of Statistics, 41(3).

[24] Maddala, G. S., & Lahiri, K. (1992). Introduction to econometrics (Vol. 2, p. 525). New York: Macmillan.